Surface state band mobility and thermopower in semiconducting bismuth nanowires

T. E. Huber,[1] A. Adeyeye,[1] A. Nikolaeva,[2,3] L. Konopko,[2,3] R. C. Johnson,[4] and M. J. Graf [4]

[1] Howard University, Washington, DC 20059

[2] Academy of Sciences, Chisinau, Moldova

[3] International Laboratory of High Magnetic Fields and Low Temperatures, Wroclaw, Poland.

[4] Department of Physics, Boston College, Chestnut Hill, MA 02467

Many thermoelectrics like Bi exhibit Rashba spin-orbit surface bands for which topological insulator behavior consisting of ultrahigh mobilities and enhanced thermopower has been predicted. Bi nanowires realize surface-only electronic transport since they become bulk insulators when they undergo the bulk semimetal-semiconductor transition as a result of quantum confinement for diameters close to 50 nm. We studied 20-, 30-, 50- and 200-nm trigonal Bi wires. Shubnikov-de Haas magnetoresistance oscillations caused by surface electrons and bulklike holes enable the determination of their densities and mobilities. Surface electrons have high mobilities exceeding 2 $m^2sec^{-1}V^{-1}$ and contribute strongly to the thermopower, dominating for temperatures $T$< 100 K. The surface thermopower is $-1.2\ T\ \mu V/K^2$, a value that is consistent with theory, raising the prospect of developing nanoscale thermoelectrics based on surface bands.

1. INTRODUCTION

In bulk solids, time reversal symmetry $[E(\vec{k},\uparrow) = E(-\vec{k},\downarrow)]$ combined with space inversion symmetry $[E(\vec{k},\uparrow) = E(-\vec{k},\uparrow)]$ depresses spin-orbit coupling. By contrast, at the crystal surface of semi-infinite surfaces of thermoelectric (TE) energy conversion materials, such as bismuth and $Bi_2Te_3$, space symmetry is lost and surface spin-orbit coupling effects are sufficiently strong that they give rise to a band of surface states. These states were observed spectroscopically with angle resolved photoemission spectroscopy (ARPES).[1] Electronic transport in surface spin-orbit bands is new in the TE field. Recently, the discovery that selected bulk TEs like $Bi_2Te_3$ and $Bi_2Se_3$ are three dimensional (3D) topological insulators (TI) created new possibilities as in TIs the surface state is protected from dissipation by time reversal symmetry ($\Re$) and therefore have exotic spintronic properties and high mobility.[2] Takahashi *et al*[3] and Ghaemi *et al*[4] presented theories of the thermopower of surface states and found that they can dominate at low temperatures. They were motivated by the report of Hor *et al*[5] of enhancements in the thermopower of $Bi_2Se_3$; however, the surface origin of the thermopower in these experiments is uncertain. This is not surprising since significant experimental hurdles exist for realizing pure surface conduction; in TIs it is observed that the likely candidates are in fact not very good bulk insulators. The surface state band mobilities in $Bi_2Te_3$ are low ($1 \times 10^4$ cm$^2$sec$^{-1}$V$^{-1}$)[6] and unobservable in $Bi_2Se_3$.[7] Here we employ Bi, an elemental semimetal characterized by low electron *n* and hole *p* densities of ($n = p = n_0 = 3 \times 10^{17}$ cm$^{-3}$), that has low intrinsic dissipation; mobilities in excess of $10^4$ m$^2$sec$^{-1}$V$^{-1}$ are attainable in Bi crystals. Bi is classified as a trivial TI as its surface states are not protected by $\Re$.[8] Also there is substantial bulk-surface state hybridization in Bi for some crystalline orientations,[9] that may circumvent TI behavior. On the other hand, surface state conduction was clearly observed in trigonal (C3//wirelength) Bi nanowires via magnetoresistance (MR).[10] In trigonal Bi wires the surface carriers were found to belong to a 3D Fermi surface (FS) with a density $N \sim \pm 1.3 \times 10^{18}$ cm$^{-3}$ (MR measurements cannot determine charges' sign) on a ~ 17 nm cylindrical sheath which amounts to a surface density $\Sigma$ of $\pm 2.2 \times 10^{12}$ cm$^{-2}$. The surface charges effective mass $m_\Sigma \sim 0.2$, in units of the electron mass m$_e$, was measured. These values are in good agreement with typical density and mass values determined via ARPES measurements ($5 \times 10^{12}$/cm$^2$) on planar surfaces.[1] Based on the bulk densities relative to $\Sigma/d$, where $d$ is the diameter, one expects surface carriers to become a majority in fine Bi wires with $d < 300$ nm at low temperatures. Furthermore, the surface-to-volume carrier density ratio is enhanced for wires on the semiconducting side of the semimetal-to-semiconductor (SMSC) transition.[11] The SMSC transition occurs when the quantum confinement energy $E_c = \hbar^2 \pi^2 / 2m^* m_e d^2$ (where $m^*$ is the bulk carrier effective mass) exceeds the electron-hole overlap energy $E_0$ and therefore, $n$ and $p$ are decreased critically below $n_0$. Because the effective masses of electrons and holes in Bi is small ($m^* \sim 0.067 <<1$), the SMSC transition occurs in wires of large diameter (50 nm for trigonal wires, $m^* = 0.067$), and since $m^* << m_\Sigma$ it should be possible to deplete the bulk carrier population in nanowires in order to reveal the properties of the surface carriers. Here we show that this strategy succeeds. Bi nanowires can be made into good bulk insulators that exhibit a surface-dominated thermopower and high mobilities.

The diffusive thermopower (mean free path (*mfp*) < wirelength) is given by: [12]

$$\alpha_\Sigma = \frac{(k_B^2 \pi^2 T / 3e)}{E_F^\Sigma} \left[ r + (d \ln N / d \ln E)_{E_F^\Sigma} \right] \quad (1)$$

Here $E_F^\Sigma$ is the surface band Fermi energy, $r = (\partial ln\tau / \partial ln E)_{E_F^\Sigma}$ where $\tau$ is the carrier lifetime and $T$ is the temperature. We assume $r \sim 0$, which is appropriate in this case as the lifetime is dominated by boundary scattering and so is energy independent. $(\partial ln N / \partial ln E)_{E_F^\Sigma} = 3/2$ in 3D. From $N = \pm 1.3 \times 10^{13} cm^{-3}$, the Fermi energy is 18 meV and we find $\alpha_\Sigma = \pm 1.2$ $T \mu V / K^2$ where the sign of the partial thermopower is related unambiguously to the sign of the charge of the carriers. In comparison, the low temperature thermopowers of electrons $\alpha_e$ and holes $\alpha_h$ in bulk Bi are found to be approximately $-1\ T\ \mu V/K^2$ and $+3\ T\ \mu V/K^2$ respectively.[13]

There have been several studies of the fabrication and of resistance $R$ and thermopower $\alpha$ of Bi wires down to diameters of around 40 nm.[14-18] The motivation was the search for nanostructured materials for efficient TE conversion. Trends are as follows: in large diameter wires ($d \sim 200$ nm), like in bulk Bi, $\alpha$ is dominated by electrons (it is type-n) because the electron mobility is two orders of magnitude larger than for holes. At intermediate diameters, ($d \sim 200$ nm), $\alpha$ has a positive bump at intermediate temperatures; an interpretation in which the mobility limitations posed by hole-boundary scattering are much less severe than those caused by electron-boundary scattering has been advanced.[18] For smaller diameters, $d < 200$ nm, $R$ saturates at low temperatures and $\alpha$ reverts to type-$n$. Previous efforts failed to identify or isolate a surface charges' thermopower, which is surprising considering that $\alpha_\Sigma$ is comparable to $\alpha_e$ and $\alpha_h$. We set out to measure the Bi nanowires $R$ and $\alpha$, from 4 K to room temperature, to explore the range of diameters (30–200 nm) that exhibited these trends in the previous studies; we were able to verify the trends with our wires. We extended this range by including 20-nm nanowire samples. Our wires, including 20-nm, demonstrate very high mobilities and display Shubnikov-de Haas (SdH) oscillations ($T < 5$ K) that allow us to measure the density of the surface and bulk species which, in turn, we correlate with $\alpha$. We will show that the trend toward negative thermopower can be associated with electrons in spin-orbit surface states. The existence of surface states also explains the observation that the resistance saturates at low $T$.

The outline of the paper is as follows. In Section 2, we examine fabrication issues and the resistance and thermopower as a function of temperature as well as the Shubnikov-de Haas data. In Section 3 we present our interpretation of the results in a multicarrier (bulklike and surface) electronic transport model. Our conclusions are presented in Section 4.

## 2. RESISTANCE AND THERMOPOWER

The samples were prepared with the template method[10] which involves fabricating a nanochannel (NC) template and filling the channels with Bi by applying pressure to the melt. The 200-nm NCs were purchased from Whatman (Shrewsbury, MA, USA). A batch of 50-nm NC was kindly provided by W. Wang of Tianjin U, China. Batches of 20-, 30-, 50- and 60-nm NCs were purchased from Synkera (Longmont, CO, USA). The error in the diameter is around 30%. A scanning electron microscope (SEM) image of the 50-nm Bi wire array is shown in Fig. 1(a). We studied the crystalline structure of the 200-nm and 30-nm wire arrays via X-ray diffraction. The rhombohedral crystal structure of bulk Bi is preserved in the wires with essentially no modification of the lattice parameters. Crystal grains (~1 μm long) were oriented with the trigonal axis along the wire length.

Figure 1(b) presents our data for the normalized resistance, the ratio of the resistance between contacts to its value at room temperature. Contacts were made using silver epoxy. The observed trends generally agree with previous measurements.[14-17] The 200-nm wires yielded bulklike metallic behavior ($dR/dT > 0$) and decreased mobility through boundary scattering, whereas the $R$ for the 50-nm, 30-nm and 20-nm wires increased monotonically with decreasing temperature, suggesting a thermally activated charge carrier density, and saturates at low temperatures. The contrast between the diameter-dependent behaviors supports the conclusion from theory regarding the SMSC transition occuring at around 50 nm.

The charge density can be measured by observing the discrete spectrum of the Landau levels (LLs) that appear because of the quantization of closed orbits in the presence of a magnetic field $B$ i.e., the SdH effect in the MR.[10] At low-$T$, less than 5 K, currents that contribute to electronic transport produce an oscillatory resistance periodic in $1/B$, and the periods (that are dependent upon the angle $\theta$ between $\vec{B}$ and the wire axis) can be mapped to obtain the FS extremal areas, and therefore the carrier densities and FS anisotropy. Figure 2(a) shows the oscillations and the LLs in the case of 50 nm wires. Figure 2(b) shows the bulklike hole $p$ and surface carrier $N$ densities; electrons are not observed in $d < 200$ nm wires. The hole band FS in large diameter wires, $d > 50$-nm, i.e 200-nm, apart from quantum confinement shifts of the Fermi energies and modest changes in the ellipsoids anisotropy, is analogous to those in bulk Bi. By decreasing the wire diameter towards 50 nm it is observed that the hole FS volume and therefore $p$, decreases critically below $n_0$. This decrease is interpreted in terms of the SMSC transition. The shape of the surface states' FS is intriguing. The anisotropy is not planar but 3D. For example in the 30-nm wires the surface and holes SdH periods are almost isotropic, the FS is nearly a sphere. Instead, in 50-nm wires both the FS of surface states and the bulklike states, within the low angular resolution of $15^0$ that has been achieved so far, is star shaped.[10] We illustrate this by contrasting the $\theta = 0$ with the $\theta = 15^0$ cases; the latter has both surface and bulklike features whereas the former case only shows bulk LL.[10] Because of the uncertainties in the FS volumes, the experimental error in *N* and *p* are large. 20-nm samples also display SdH oscillations from surface and bulk bands and in this case the FS are relatively fairly isotropic. The SMSC transition argument would lead us to expect that if the 50 nm wires exhibit a strong decrease in electron and hole density, then those with *d* < 50 nm, such as 30-nm and 20-nm, should exhibit equal or larger decreases but this not observed. *p* has a minimum for *d* ~ 50 nm and for decreasing *d*, for 30-nm and 20-nm wires, *p* is substantial. The *N/p* ratio is ~8 in 30-nm wires. In comparison ARPES study of 27 nm-thick films show a surface-to-bulk density ratio of 100, presumably due to hybridization of the surface and bulklike carrier populations.[9,19] This low level of discrepancy regarding surface-bulk ratios is gratifying considering the different geometries and crystal surfaces orientations.

Employment of the Lifshitz-Kosevich expressions for the field and temperature dependencies of the SdH oscillation amplitude in 3D *FS* allows for the determination of the carriers' effective masses $m^{eff}$ and mobility $\mu$.[20]

$$MR(T,B) = \left(\frac{\hbar\omega_c}{2\varepsilon_F}\right)\frac{2\pi^2 k_B T/\hbar\omega_c}{sinh(2\pi^2 k_B T/\hbar\omega_c)} exp(-2\pi^2 k_B T_D/\hbar\omega_c) \qquad (2)$$

where $\hbar\omega_c = 2\mu_B B/m^{eff}$, where $m^{eff}$ is either $m_\Sigma$ or $m^*$ and $T_D$ is the Dingle temperature. $\tau$, the semi-classical transport collision time, is $\tau > \tau_D = \hbar/k_B T_D$.

$\mu > e\tau_D/m^{eff}m_e$ ; estimates are a lower limit since the actual relaxation time $\tau$ is longer than the broadening embedded in $\tau_D$.

By fitting equation 2 to the 2 K and 5 K 30-nm wires surface carriers SdH amplitudes, that have $m_\Sigma$ =0.21 ± 0.03, we find $T_D$=3 K. Therefore we estimate $\tau_D$ ~2× $10^{-12}$ sec and $\mu_\Sigma \geq 1.7$ m$^2$ sec$^{-1}$ V$^{-1}$. Also, we find $\mu_h \geq 2.0$ m$^2$ sec$^{-1}$ V$^{-1}$. Aside from the effective mass, the surface nature is inferred from the near diameter independence of $n_S$. The carrier penetration length into the sample, $\lambda$, can be estimated from the SdH data because, to show LL, the spatial range of the surface states has to be larger than the Larmor radius $r_L$. Since $r_L = \frac{m_\Sigma m_e V_F}{|e|B}$, where the 3D Fermi velocity $V_F = \sqrt{2E_F / m_\Sigma m_e} \sim 5 \times 10^4$ m/sec, the lower end of the range of magnetic fields where the surface LLs are observed marks the spatial extent of the sheath of charges. (See Fig. 2(a), inset). Because the surface LLs become unobservable for $B = 5$ Tesla (T), when $r_L$ is roughly 17 nm, this is our estimate of $\lambda$. Clearly, the surface states fill a very substantial fraction of the wire, especially in the 20-nm case. The charge density per unit area $N$ is roughly $n_S\lambda$, where $n_S$ can be estimated from the FS volume that is estimated from the SdH measurements. $N$, which is observed to be roughly the same for wires of diameters between 30 nm and 200 nm, is found to be $2.2 \times 10^{12}$/cm$^2$.

The semiconductor-like behavior of $R(T)$ is most apparent in the 20-nm wires. The wires' conductance per unit length is $G(T) = G_{BL} + G_\Sigma$, where the first and second terms are the bulklike and surface contributions, respectively. Here, $G_{BL}(T) = (\pi d^2/4)\sigma_{BL}(T)$, where the term $(\pi d^2/4)$ accounts for the wire cross-sectional area, and $\sigma_{BL}(T) = en_{BL}(T)\mu_{BL}(T)$. Also, $G_\Sigma = \pi d e \Sigma \mu_\Sigma$, where the term $\pi d$ is the perimeter. $\mu_{BL}$ and $\mu_\Sigma$ are the bulklike and surface states' mobility, respectively. $e$ is the charge of the free electron. At low temperatures, $R$ reaches saturation instead of the exponential $T$ dependence characteristic of semiconductors. Saturation has been attributed to uncontrolled impurities.[15,16] The effect can be caused by surface band conduction also. A close low bound estimate of the $G_\Sigma$ (4 K), and the resistance square $R\square$ $=\pi d / G_\Sigma (4K)$ can be obtained as follows. We consider that the wire resistance at room temperature $R(300 K)$ is less than $\rho_0 \dfrac{L}{(\pi/4)d^2}$, where $\rho_0$ is the 300 K resistivity;[13] this is a close inequality because finite size effects are small and positive at 300 K. Also we have measured the normalized resistance, the resistance ratios $R(4 K)/R(300 K)$ (Fig. 1.b). Therefore, we find that upper limits of $R_\square$ are 150 Ω, 180 Ω and 430 Ω for 50, 30 and 20 nm respectively. These values are not inconsistent with other determinations of the resistivity for surface bands; for example. Hirahara *et al* [19] found $R_\square$ of ultrathin Bi films to be 670 Ω. Since Σ can be estimated for our wires as $N/\lambda$, the mobility of the surface states $\mu_\Sigma$ is found to be 2.1 $m^2V^{-1}sec^{-1}$, 1.7 $m^2V^{-1}sec^{-1}$ and *0.7 $m^2V^{-1}sec^{-1}$* for 50, 30 and 20 nm respectively. These estimates are close to whose obtained by the SdH method. The thermally activated dependence at high temperatures can be explained if $n_{BL}(T) \sim exp(-\varepsilon/T)$. Following the procedures of Choi *et al*[14] [see Eq. (1) of this paper], $\varepsilon$ is found to be 40 ± 4 meV and is interpreted as the energy gap between the electron and hole bands in the core of the 20-nm wires. Figure 1(b) shows the fit. We assume that $\mu_\Sigma$ is temperature-independent. The positive and high value of $\varepsilon$ does indeed indicate that the band overlap decreases substantially below the value for bulk Bi (38 meV) to become a gap in the small diameter wires because of quantum confinement further supporting the SdH evidence for the onset of the SMSC transition near 50 nm. Surface transport dominates over bulk transport up to about 100 K for 50- and 20-nm and to about 50 K for 30 nm.

Figure 3(a) shows the thermopower $\alpha$ =(Thermoelectric voltage)/($T_{hot}$ - $T_{cold}$) of the samples; the temperature difference was ≤ 3 K. Figure 3(b) illustrates the experimental setup for the thermopower measurements. This device was mounted in a closed cycle refrigerator with a temperature range from 4 K to room temperature. Wire array samples, which were a fraction of a mm thick and with a surface area ~1mm², were contacted through the anvil pieces, using indium (In) foil as an interface material. The heater allowed us to create a temperature difference between the two electrodes and, therefore, between the two ends of the wires. The observed trends of the thermopower agree with previous studies of Bi wire arrays.

3. ELECTRONIC TRANSPORT. BULKLIKE AND SURFACE BAND CONTRIBUTIONS

Our interpretation of the data incorporates surface bands. $\alpha$ is the weighed average of the partial thermopowers of surface, $\alpha_\Sigma$, and bulklike, $\alpha_{BL}$, bands:

$$\alpha = \frac{G_\Sigma \alpha_\Sigma + G_{BL} \alpha_{BL}}{G_\Sigma + G_{BL}} \tag{3}$$

$\alpha_{BL}$ is the weighted average of the thermopower of bulklike electrons and holes where the weight factors are the electron, $G_e$, and hole, $G_h$, conductances. $G_{BL} = G_e + G_h$. At low temperatures, the thermopower of the 50-nm samples should display the strongest evidence of thermopower by surface states because electrons are not observed ($n=0$) and $p$ is small ($p<p_0$), therefore the bulklike conductivity can be neglected and the surface charges predominate over the holes. Indeed, the decrease of the thermopower for $T < 50$ K is linear in $T$, with a slope of $-1.25$ μV/K² and this value is in very good agreement with the thermopower estimate from Eq. 1 using $\Sigma = N/\lambda$ from Fig. 2(b) that is $\alpha_\Sigma = -1.2$ μV/K². Since we find that the surface thermopower is negative, the surface charges are electrons. Figure 3(a) shows that, at low-$T$, $T <$ 20 K, the thermopower of the 30-nm and 20-nm wires can be approximated by -0.08 $T$ μV/K² and +0.04 $T$ μV/K², respectively. These thermopowers are interpreted in terms of Equation 3. The holes' Fermi energy is $E_F^h \approx p^{2/3}$, where $p=p_0$ for $E_F^h = 11$ meV.[10] Since $p = 1.3 \times 10^{17}$ cm$^{-3}$, the $E_F^h$ of the holes in the 30-nm and 20-nm wires is 2.2 meV. Holes have thermopower $\alpha_h = \left(\frac{k_B^2 \pi^2 T}{3e}\right)\left[\frac{1}{E_F^h}\right]$ and we find that $\alpha_h = 17\, T\, \mu V/K^2$. Therefore, the ratios $G_h/G_\Sigma$ are $0.01 \pm 0.01$ for 50-nm and $0.08 \pm 0.01$ for 30- and 20-nm wires. From $n_h$ and $p$ [Fig. 2(b)] we find $\mu_h = 1.4\, m^2 V^{-1} sec^{-1}$ for both 30- and 20-nm wires.

In the 50-nm wires, the low-$T$ line is followed by a positive bump that peaks at -5 μV/K at $T \sim$ 120 K. This bump is similar to the bump that is observed in the 200-nm wires at 40 K, as well as to the bumps that have also been observed by Boukai *et al*[17] in 60-nm wires. An interpretation[18] is that this positive bump arises because the interplay between electrons and holes considering that the mobility limitations posed by hole-boundary scattering are much less severe than those caused by electron-boundary scattering. Therefore, at $T \sim$ 120 K, surface charges are together with both bulklike electrons and holes. In the 30-nm and 20-nm case the bumps arise at much lower temperatures (~20 K). What makes these peaks interesting is their similarity to, and possible commonality with, those observed by Hor *et al*[5] at the same temperatures.

Because the 50-nm wire samples displayed the strongest evidence of thermopower by surface states and by holes, we attempted to modify the surface thermopower term by doping. We investigated the effect of doping the wires with tin (Sn; electron acceptor) and tellurium (Te; electron donor), which change the position of the Fermi level and affect the balance of densities between electrons and holes.[21] We used templates from a single batch prepared in an identical fashion and observed that the bump at intermediate temperatures, $T \sim$ 120 K, which was positive in pure Bi, becames less or more pronounced for Te- and Sn-doped wires, respectively. This is expected from the shifting electron-hole density balance considering our assignment of this peak as a bulklike electron-hole interplay. Attempts to manipulate the surface thermopower contribution by doping were unsuccessful; we observed that the linear term at intermediate temperatures that we assigned to the surface states is unchanged, within 10%, even under conditions of strong doping. Overall, these tests confirm the interpretation of the surface and bulklike features presented in the previous paragraphs and hint that the surface term contributes, not only at low-$T$, but also at moderate temperatures.

## 4. CONCLUSIONS

In conclusion, while the theory of topological insulators is generating a lot of interest substantial hurdles exist for realizing the predictions since the candidates are poor bulk insulators. Our proposal for promoting bulk insulator behavior is to shape the TEs into nanowires that become semiconductors as a result of quantum confinement. We tested Bi nanowires that show strong quantum confinement effects due to small effective mass. 50-nm wires feature almost pure surface conduction. Bi is not a true topological insulator because the time reversal symmetry is not conserved. Still, we have achieved mobilities of over 2 m$^2$sec$^{-1}$V$^{-1}$ with a density $2.2 \times 10^{12}$ cm$^{-2}$. This mobility is 2/3 of the values that are found for unsuspended graphene with significantly less charge densities.[22] Our mobility values are twice those found by Qu *et al*[6] for Bi$_2$Te$_3$ surface bands. Our high value of surface mobility appear to be related to the special conditions in 50-nm wires since it is significantly less for 30- and 20-nm wires. Also with the nanowire approach, we have shown that the surface contributes strongly to the thermopower, dominating for temperatures $T$< 100 K. Surface thermopower is $-1.2\ T$ μV/K$^2$, a large value that is consistent with theory, raising the prospect of developing nanoscale thermoelectrics based on surface bands. These findings indicate that the approach of experimenting with low dimensional structures of topological insulators and leveraging quantum confinement effects to enlarge surface or bulk gaps, may lead to more successful TE materials. Considerable advances are being made in this direction in theoretical studies as well.[23]

This material is based upon work supported by the National Science Foundation (NSF) under Grant No. NSF-DMR-0839955 and NSF-DMR-0611595 and by the U.S. Army Research Office Materials Science Division under Grant No. W911NF-09-1-05-29. We also acknowledge support by the Boeing Corporation and Swiss National Science Foundation SCOPES.

# REFERENCES

*Corresponding author: T.E. Huber. Address: Howard University, Washington, DC 20059, USA. Tel: 202 806 6768. *Email address*:  **thuber@howard.edu**.

**FIGURE CAPTIONS**

**FIG. 1.** (a) SEM image of the top of a 50-nm Bi nanowire array. Light spots represent nanowires. Electron energy is 10 keV, and magnification is 40,000. (b) Normalized resistance of arrays of 200-, 50-, 30-, and 20-nm Bi nanowires as indicated. The dashed line is the fit using our two-channel (surface and bulklike hole) model. The bulklike electron-hole gap is 40 meV.

**FIG. 2.** (a) Solid line: 1.5 K-MR of 50-nm Bi wires for two orientations $\theta$ of applied field. Dotted-solid line: 1.5 K-MR of 20-nm Bi wires for $\theta = 0$. Bulklike (BL) and surface extrema associated with Landau levels are shown with vertical lines. Inset: 50-nm 1/B-sequences. (b) Low-$T$ hole $p$ and surface charge density $N$ as a function of inverse diameter for bulk Bi and for Bi wires. Diameter measurements have large (~30%) errors. The semimetal-semiconductor (SMSC) transition is indicated. Inset: Representation of the surface charges (solid gray) and bulklike charges in (pattern) the two-channel model presented in this paper. $\lambda$ is the thickness of the sheath of surface charge.

**FIG. 3.**(a) Thermopower of 200-, 50-, 30-, and 20-nm Bi nanowires as indicated. 40-nm data from Lin, Rabin, Cronin, Ying and Dresselhaus[16] is also shown. The dashed line in the 50-nm data is a linear fit. (b) Inset: Experimental set-up for thermopower.

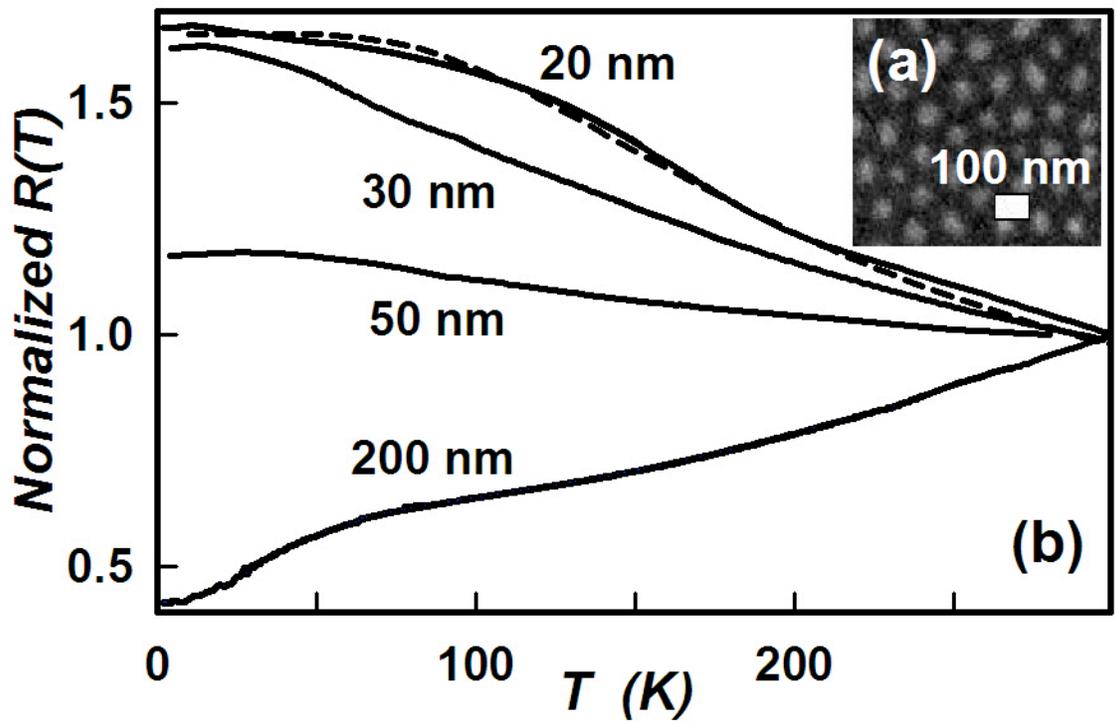

Fig. 1(a) and (b). Huber *et al*. 2011.

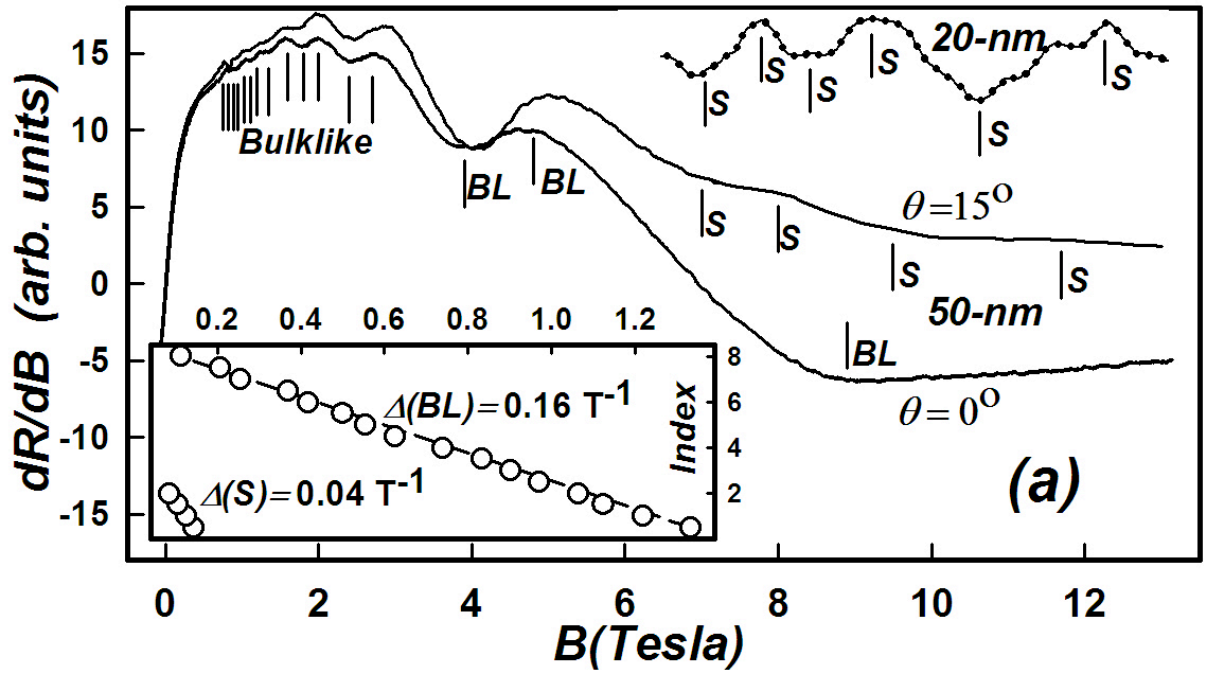

Fig. 2(a). Huber *et al.*. 2011.

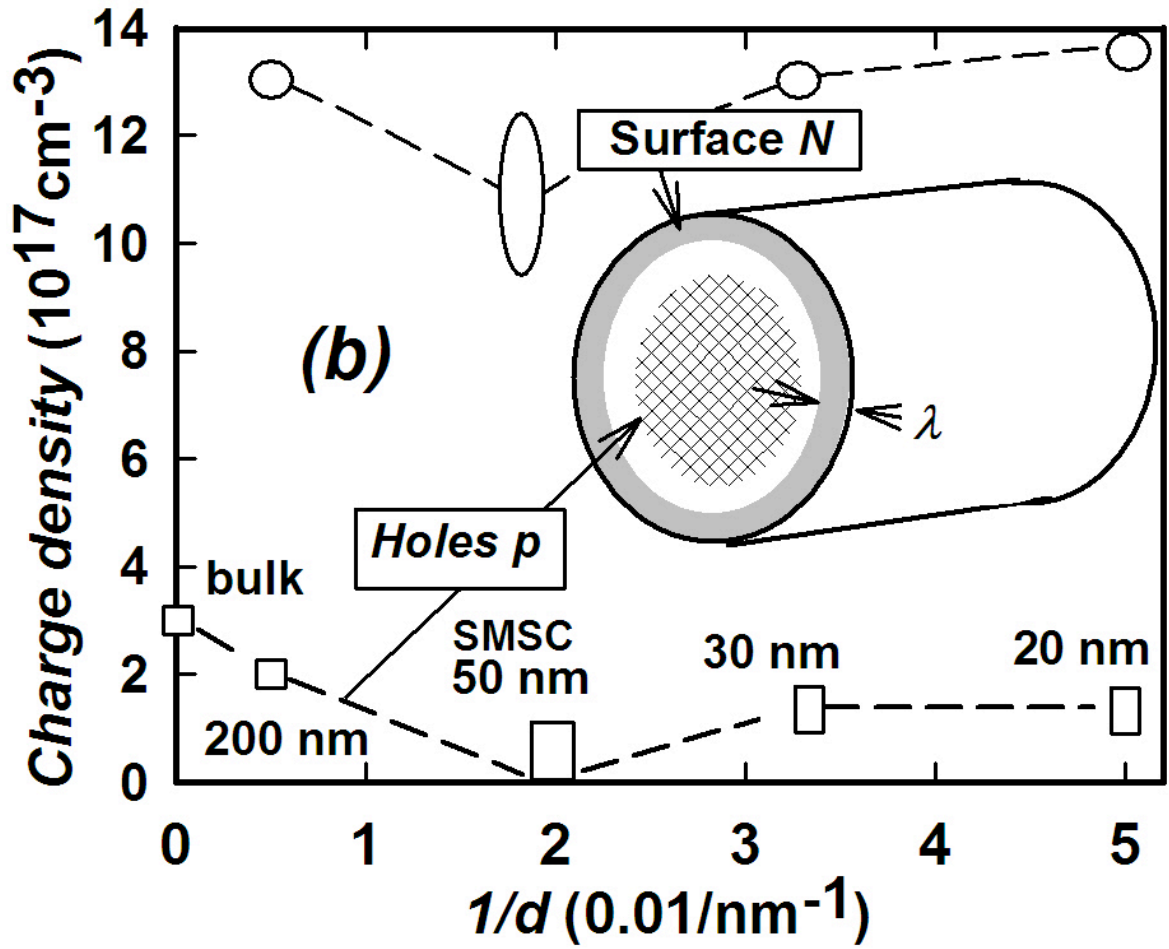

Fig. 2(b). Huber *et al.*. 2011.

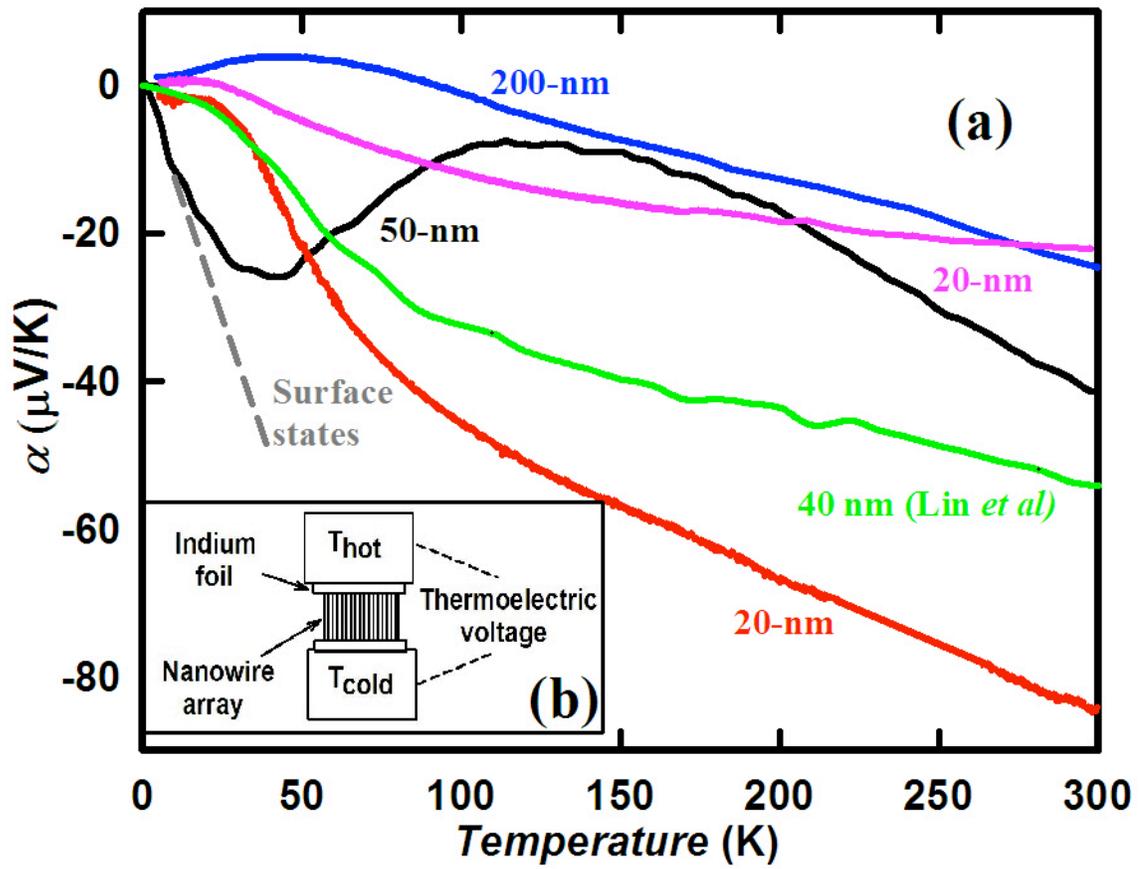

Fig. 3.(a) and (b). Huber *et al.*. 2011.